\documentclass[12pt]{article}
\usepackage{graphicx}
\usepackage{amsfonts}
\usepackage{amssymb,amsmath}
\usepackage{latexsym}
\usepackage{color}
\usepackage{bm}
\usepackage{cite}
\input{colordvi.tex}

\newcommand{\bi}{\begin{itemize}}
\newcommand{\ei}{\end{itemize}}
\newcommand{\beq}{\begin{equation}}
\newcommand{\eeq}{\end{equation}}
\newcommand{\bea}{\begin{eqnarray}}
\newcommand{\eea}{\end{eqnarray}}

\def\curv{\sigma}
\def\mod{\chi}

\def\tf{\tilde f_{\rm NL}}
\def\tg{\tilde g_{\rm NL}}

\def\d{\delta}

\begin{document}

\begin{titlepage}

\begin{center}

{\Large \bf  
Density Perturbations  \\ \vskip 3pt
from Modulated Decay of the Curvaton}

\vskip .45in

{\large
David Langlois$^1$ and
Tomo Takahashi$^2$
}

\vskip .45in

{\em
$^1$APC (CNRS-Universit\'e Paris 7), 
10, rue Alice Domon et L\'eonie Duquet, 75205 \\ Paris Cedex 13, France \\
\vspace{0.2cm}
$^2$Department of Physics, Saga University, Saga 840-8502, Japan 
  }

\end{center}

\vskip .4in

\begin{abstract}
  We study density perturbations, including their non-Gaussianity, in
  models in which the decay rate of the curvaton depends on another
  light scalar field, denoted the modulaton.  Although this model
  shares some similarities with the standard curvaton and modulated
  reheating scenarios, it exhibits interesting predictions for $f_{\rm
    NL}$ and $g_{\rm NL}$ that are specific to this model. We also
  discuss the possibility that both modulaton and curvaton
  fluctuations contribute to the final curvature perturbation.  Our
  results naturally include the standard curvaton and modulated
  reheating scenarios as specific limits and are thus useful to
  present a unified treatment of these models and their variants.

\end{abstract}

\end{titlepage}

\clearpage

\setcounter{page}{1}
\setcounter{footnote}{0}

\bigskip

\section{Introduction}

Cosmological observations are increasingly precise, providing us with
a lot of information on the origin of cosmic structure, i.e. on
primordial density fluctuations.  Although the fluctuations of the
inflaton field are considered as the main candidate for their origin,
other possibilities have also been investigated (see
e.g. \cite{Langlois:2010xc} for introductory lectures), especially in
the light of recent constraints on primordial non-Gaussianity.  The
degree of non-Gaussianity in primordial density fluctuations is
usually characterized by the non-linearity parameter $f_{\rm NL}$
which represents the amplitude of the bispectrum.  In the case of
standard slow-roll single-field inflation, $f_{\rm NL}$ is predicted
to be too small to be observable.  On the other hand, the present
constraints on $f_{\rm NL}$ for local type non-Gaussianity obtained
from cosmic microwave background (CMB) and large scale structure are
respectively $f_{\rm NL}^{\rm (local)} = 37.2 \pm 19.9$ (1$\sigma$
C.L.) from WMAP9 \cite{Bennett:2012fp} and $f_{\rm NL}^{\rm (local)} =
62 \pm 27$ (1$\sigma$ C.L.)  from NRAO VLA Sky Survey
\cite{Xia:2010yu}, which may give some hints that the value of $f_{\rm
  NL}$ is away from zero.

In this context, other candidates for primordial fluctuations, in
particular those giving significant $f_{\rm NL}$, have been
extensively discussed such as the curvaton model
\cite{Enqvist:2001zp,Lyth:2001nq,Moroi:2001ct}, modulated reheating
scenario \cite{Dvali:2003em,Kofman:2003nx}, inhomogeneous end of
hybrid inflation
\cite{Bernardeau:2002jf,Bernardeau:2004zz,Lyth:2005qk,Salem:2005nd},
modulated trapping \cite{Langlois:2009jp,Battefeld:2011yj} and so
on. Even if we limit ourselves to the curvaton and modulated reheating
scenarios, various extensions of them have been proposed and studied,
for example mixed inflaton-curvaton model
\cite{Langlois:2004nn,Lazarides:2004we,Moroi:2005kz,Moroi:2005np,Ichikawa:2008iq},
mixed inflaton-modulated reheating model \cite{Ichikawa:2008ne},
multi-curvaton \cite{Choi:2007fya,Assadullahi:2007uw}, modulated
curvaton \cite{Suyama:2010uj,Choi:2012te} and so on\footnote{
  In most works, the curvaton potential is assumed to have a quadratic
  form, however, other types of the potential have also been discussed
  such as curvaton with self-interaction
  \cite{Enqvist:2005pg,Enqvist:2008gk,Enqvist:2009zf,Enqvist:2009eq,Enqvist:2009ww}
  and pseudo-Nambu-Goldstone curvaton
  \cite{Dimopoulos:2003az,Kawasaki:2008mc,Chingangbam:2009xi,Kawasaki:2011pd}.
}. In most of those scenarios, a light scalar field (degree of
freedom) other than the inflaton is involved in some way, and the
final values of density perturbations depend on how the initial
fluctuations are converted to the final ones during the evolution of
the early Universe. This consideration provides a strong motivation to
treat models involving curvatons and/or modulatons in a unified
formalism (see \cite{Langlois_Takahashi}).  In the present work, we
wish to focus our attention on a scenario that interpolates between
the modulated reheating and the curvaton models: that of the modulated
curvaton decay, in which the decay of the curvaton field is modulated
by the dependence of its decay rate $\Gamma$ on another fluctuating
scalar field, which we will call the modulaton\footnote{
  Our scenario differs from the so-called modulated curvaton model,
  discussed in \cite{Suyama:2010uj,Choi:2012te}, where the curvaton
  first plays the role of a modulaton during the decay of the
  inflaton, then decays at some later time as in the usual curvaton
  model.
}.

In this paper, we focus on this new mechanism of generating primordial
density fluctuations and derive its predictions, paying particular
attention to non-Gaussianities by computing the bispectrum and
trispectrum.  Although the model we propose here is, in some sense, a
straightforward extension of the curvaton and modulated reheating
scenarios, we find that it leads to a rich phenomenology, with
interesting observational implications for primordial non-Gaussianity.
We also provide general formulas, which could be applied to other
similar types of scenarios (see \cite{Langlois_Takahashi} for a
systematic approach, including isocurvature perturbations).

The structure of this paper is as follows. In the next section, we
derive a general expression describing the final curvature
perturbation, up to third order in the perturbations.  Focussing then,
in Section~\ref{sec:modulaton_dom}, on the modulaton fluctuations, we
analyse the density perturbation and compute the non-linear parameters
such as $f_{\rm NL}$ and $g_{\rm NL}$.  In Section~\ref{sec:hybrid},
we investigate models in which the curvaton fluctuations also
contribute to the observed perturbations in addition to those from the
modulaton.  The final section contains a summary of this paper.

\section{Computing the post-decay perturbation}
\label{sec:general}
The present scenario relies on the presence of three scalar fields: an
inflaton field $\phi$ which drives the inflationary expansion; a
curvaton (or modulus) field $\curv$, with an energy density negligible
during inflation, which, well after inflation, oscillates at the
bottom of its potential before decaying; and, finally, a modulaton
field $\mod$, which is light during inflation and thus acquires
fluctuations from the amplification of quantum fluctuations.  The
crucial assumption here is that the decay rate of the curvaton $\curv$
depends on the modulaton $\chi$. Therefore, fluctuations of $\mod$
directly lead to a varying decay rate and eventually produce density
fluctuations.

In practice, we will not need any detail about the inflaton field. Its
role will be simply to drive inflation so that the modulaton field can
acquire some fluctuations. In our scenario, $\sigma$ can be either
light during inflation ($m_\sigma\ll H$), in which case it will also
acquire some fluctuations, or be massive ($m_\sigma\gg H$) in which
case its fluctuations are suppressed. Strictly speaking, the curvaton
scenario assumes a light field during inflation, but there exist
models where the additional scalar fields, such as moduli, are not
necessary light during inflation. Whereas the original curvaton
scenario would not apply to scalar fields of this type, our model
does.  In the following, although $\sigma$ is a scalar field, we will
be interested in the cosmological phase where it oscillates at the
bottom of its potential and can be effectively described as a fluid,
which is pressureless if the potential is quadratic.  Note that our
formalism also applies to the decay of the inflaton oscillating in a
quadratic potential at the end of inflation, if $\sigma$ is simply
replaced by the inflaton $\phi$.  Our formalism thus includes
automatically the modulated reheating scenario.

For each fluid characterized by an equation of state parameter
$w_i\equiv P_i/\rho_i$, which is assumed here to be constant, it is
convenient to introduce the non-linear curvature perturbation
$\zeta_i$ \cite{Lyth:2004gb} (see also
\cite{Langlois:2005ii,Langlois:2005qp,Langlois:2010vx} for a covariant
definition)
\begin{equation}
\label{eq:zeta_i}
\zeta_i = \delta N + \frac{1}{3(1+w_i)} \ln \left( \frac{\rho_i ( t, \vec{x})}{\bar{\rho}_i(t)} \right), 
\end{equation}
where $\delta N$ denotes the local perturbation of the number of
e-folds and a barred quantity must be understood as homogeneous.  From
the above formula, the nonlinear energy density of the species $i$ can
be written locally as
\begin{equation}
\label{eq:rho_i}
\rho_i (t, \vec{x}) = \bar{\rho}_i(t) e^{3 (1+ w_i) (\zeta_i - \delta N)}.
\end{equation}
In our case, we will consider only two  species: radiation
($w_r=1/3$)
and 
 the curvaton field, treated as a pressureless fluid
($w_\curv=0$).

Using the instantaneous decay approximation, the value of the Hubble
parameter at the decay of the curvaton $\sigma$ (or, alternatively, of
the inflaton to describe inhomogeneous reheating) is given by
\beq
H=\Gamma (\mod),
\eeq
where the decay rate $\Gamma$ is a function of the modulaton
$\mod$. Because of the modulaton fluctuations $\delta\mod =H/(2\pi)$
generated during inflation, the decay hypersurface characterized by
the above relation is {\it inhomogeneous}.  Using Friedmann's
equations\footnote{Note that we are implicitly using the separate
  Universe approach where distinct regions are described by FLRW
  universes. This is justified by the fact that the perturbations we
  are interested in are on super-Hubble scales at the time of the
  decay.}, this implies for the local energy density
\begin{equation}
\label{eq:rho_tot_D_1}
\frac{\rho_{\rm tot} (t_D,\vec{x})}{\bar{\rho}_{\rm tot} (\bar{t}_D)} 
=
\frac{H^2 (t_D,\vec{x})}{\bar{H}^2 ({t}_D)} 
=
\frac{\Gamma^2 (t_D,\vec{x})}{\bar{\Gamma}^2 (\bar{t}_D)},
\end{equation}
where  $t_D(\vec{x})$ represents  the local decay time. Substituting 
\beq
\rho_{\rm tot} (t_D,\vec{x})=\sum_i\rho_i (t_D, \vec{x}) =\sum_i \bar{\rho}_i(\bar{t}_D)\,  e^{3 (1+ w_i) (\zeta_i - \delta N_D)}
\eeq
in the relation \eqref{eq:rho_tot_D_1}, both for the matter contents just before and just after decay,
 we find 
\begin{equation}
\label{eq:energy_density_eq}
(1- \Omega_\curv) \, e^{4  (\zeta_{r} - \delta N_D)}+\Omega_{\curv} \, e^{3  (\zeta_{\curv} - \delta N_D)}
= 
 \left( 1+\d_\Gamma \right)^2 
=
e^{4(\zeta - \delta N_D)},
\end{equation} 
where we have introduced the curvaton fraction of the total energy
density (just before the decay) $\Omega_\curv\equiv
\bar{\rho}_{\curv}/\bar{\rho}_{\rm tot}$, as well as the (nonlinear)
relative fluctuations of the decay rate
\beq
\delta_\Gamma\equiv \frac{\Gamma(t_D,\vec{x})}{\bar{\Gamma}(\bar{t}_D)}-1\,.
\eeq
The first equality in \eqref{eq:energy_density_eq} gives us the
expression of $\delta N_D$ as a function of the two pre-decay
curvature perturbations $\zeta_r$ and $\zeta_\curv$. And the second
equality in \eqref{eq:energy_density_eq} yields the expression of the
post-decay curvature perturbation $\zeta$ (carried by the
only-remaining radiation fluid) as a function of $\delta N_D$, namely
\beq
\label{zeta}
\zeta=\d N_D+\frac12 \ln(1+\d_\Gamma)\,.
\eeq
There is no general nonlinear expression for $\delta N_D$ given in
terms of $\zeta_r$ and $\zeta_\curv$, but by expanding the first
equality of \eqref{eq:energy_density_eq} order by order in the
perturbations, one can iteratively obtain an explicit expression for
$\delta N_D$ valid up to any order. Computing $\delta N_D$ up to third
order with this method and substituting in (\ref{zeta}), we finally
get the following expression for the post-decay curvature
perturbation:
\begin{eqnarray}
\label{zeta_3}
\zeta&=& \zeta_r-\frac{r}{6}\delta _{\Gamma }+\frac{r}{3} S 
-\frac{r}{72}\left[\left(r^2+2 r-9\right) \delta _{\Gamma }^2+4 \left(r^2+2
   r-3\right) S^2-4 \left(r^2+2 r-3\right) S \delta _{\Gamma }\right]
   \notag \\
   &&+\frac{r}{1296} \left[\left(-3 r^4-10 r^3+22 r^2+54 r-135\right) \delta _{\Gamma }^3+8
   \left(3 r^4+10 r^3-4 r^2-18 r+9\right) S^3
   \right.
   \notag \\
   &&\left. \ 
   -12 \left(3 r^4+10 r^3-4 r^2-18
   r+9\right) S^2 \delta _{\Gamma }+6 \left(3 r^4+10 r^3-10 r^2-30 r+27\right) S
   \delta _{\Gamma }^2\right], \notag \\
\end{eqnarray}
where we have introduced, for convenience, the curvaton isocurvature
perturbation
\beq 
S\equiv 3(\zeta_\sigma-\zeta_r)
\eeq
and the parameter $r$, defined by
\begin{equation}
r \equiv  \left. \frac{3 \bar{\rho}_\sigma}{4 \bar{\rho}_r + 3 \bar{\rho}_\sigma} \right|_{D}=\frac{3\,\Omega_\sigma}{4-\Omega_\sigma}\,.
\end{equation}
Although isocurvature fluctuations can also be generated in principle,
we restrict our analysis to adiabatic perturbations in the present
work (see \cite{Langlois_Takahashi} for an analysis including
isocurvature modes).

Note that the perturbations $\zeta_r$, $\zeta_\curv$ and
$\delta_\Gamma$ are related to the fluctuations of the inflaton,
curvaton and modulaton, via the expressions\footnote{We include only
  the linear term for the inflaton fluctuations, since their
  non-Gaussianity can be neglected in the simplest models.}  
\beq
\label{expansions}
\zeta_r=\frac{H}{\dot\phi}\delta\phi, 
\quad 
S=\frac{2\d\curv}{\curv} -  \frac{\d\curv^2}{\curv^2} + \frac23 \frac{\d\curv^3}{\curv^3}, 
\quad 
\d_\Gamma=
 \frac{\Gamma'}{\Gamma}\d\mod+\frac{\Gamma''}{2 \Gamma}\d\mod^2+\frac{\Gamma^{'''}}{6 \Gamma}\d\mod^3,
\eeq
where a prime denotes the derivatives with respect to
$\mod$. Substituting the above expressions in (\ref{zeta_3}) would
thus give the curvature perturbation $\zeta$ as a function of the
fluctuations $\delta\phi$, $\delta\curv$ and $\delta\mod$.

Once the curvature perturbation has been computed, here up to third
order, it is useful, in order to confront the model with observations,
to calculate the power spectrum~$P_\zeta$, bispectrum~$B_\zeta$ and
trispectrum~$T_\zeta$. They correspond, respectively, to the 2-point,
3-point and 4-point correlation functions in Fourier space:
\begin{eqnarray}
\langle \zeta_{{\bm k}_1} \zeta_{{\bm k}_2} \rangle 
&=&
 (2 \pi)^3 \delta ({\bm k}_1+{\bm k}_2)  P_\zeta(k_1)\,,  \\
\langle \zeta_{{\bm k}_1} \zeta_{{\bm k}_2} \zeta_{{\bm k}_3} \rangle 
&=&
 (2 \pi)^3 \delta ({\bm k}_1+{\bm k}_2+{\bm k}_3) B_\zeta(k_1, k_2, k_3)\,,  \\
\langle \zeta_{{\bm k}_1} \zeta_{{\bm k}_2} \zeta_{{\bm k}_3}  \zeta_{{\bm k}_4} \rangle 
&=&
 (2 \pi)^3 \delta ({\bm k}_1+{\bm k}_2+{\bm k}_3 + {\bm k}_4 ) T_\zeta(k_1, k_2, k_3, k_4)\,.
 \end{eqnarray}
 In the case of local non-Gaussianity, it is convenient to express the
 bispectrum and trispectrum in terms of the power spectrum and to
 introduce the so-called non-linearity parameters $f_{\rm NL}$ for the
 bispectrum, $\tau_{\rm NL}$ and $g_{\rm NL}$ for the trispectrum:
\begin{eqnarray}
\label{eq:def_f_NL}
B_\zeta (k_1,k_2,k_3)
&=&
\frac{6}{5} f_{\rm NL} 
\left( 
P_\zeta (k_1) P_\zeta (k_2) 
+ P_\zeta (k_2) P_\zeta (k_3) 
+ P_\zeta (k_3) P_\zeta (k_1)
\right), \\
T_\zeta (k_1,k_2,k_3,k_4)
&=&
\tau_{\rm NL} \left( 
P_\zeta(k_{13}) P_\zeta (k_3) P_\zeta (k_4)+11~{\rm perms.} 
\right) \nonumber \\
&&
+ \frac{54}{25} g_{\rm NL} \left( P_\zeta (k_2) P_\zeta (k_3) P_\zeta (k_4)
+3~{\rm perms.} \right).
\label{eq:def_tau_g_NL} 
\end{eqnarray}

Quite generically, if the curvature perturbation can be written in the form
\begin{equation}
\label{N_exp}
\zeta = 
N_a \delta \varphi^a + \frac12 N_{ab} \, \delta \varphi^a \delta \varphi^b  
+ \frac16 N_{abc}  \, \delta \varphi^a \delta \varphi^b \delta \varphi^c
+ \cdots, 
\end{equation}
where the $\varphi^a$ denotes any number of light scalar fields,
labelled by the index $a$, with statistical independent fluctuations
generated during inflation\footnote{We implicitly assume, for
  simplicity, slow-roll inflation with standard kinetic terms for all
  scalar fields, so that their fluctuations are approximately
  Gaussian.},
\beq
\langle \d\varphi^a_{{\bm k}_1} \d\varphi^b_{{\bm k}_2} \rangle=  (2 \pi)^3 \delta ({\bm k}_1+{\bm k}_2)  P_*(k_1) \delta^{ab}, \quad P_*(k_1)=\frac{4\pi^2}{k_1^3}{\cal P}_*, \quad {\cal P}_*=\left(\frac{H_*}{2\pi}\right)^2\,,
\eeq
the power spectrum is given by 
\beq
\label{power}
P_\zeta= N_a N^a\,  P_*\,,
\eeq
and the non-linearity parameters by the simple expressions
\cite{Lyth:2005fi,Alabidi:2005qi,Byrnes:2006vq}
\begin{eqnarray}
\tf\equiv \frac{6}{5}f_{\rm NL}
&=&
 \frac{N_a N_b N^{ab}}{ {(N_c N^c)}^2 }, 
\label{eq:f_NL_N} \\
\tau_{\rm NL}
&=&
\frac{N_{ab} N^{ac} N^b N_c}{ {(N_d N^d)}^3 }, 
\label{eq:tau_NL_N} \\
\tg\equiv \frac{54}{25} g_{\rm NL}
&=&
\frac{N_{abc} N^a N^b N^c}{ {(N_d N^d)}^3 }\,,
\label{eq:g_NL_N}
\end{eqnarray}
where we have used the Kronecker symbols to raise the scalar field
indices, e.g. $N^a\equiv \delta^{ab}N_b$.

Equipped with the general formalism presented in this section, we now
discuss in more details various scenarios based on the modulated decay
of the curvaton in the following sections.

\section{Modulaton dominated case}
\label{sec:modulaton_dom}
In this section, we restrict ourselves to the simplest case where only
the modulaton fluctuations are relevant, i.e. we neglect the
fluctuations of the curvaton and of the pre-decay radiation fluid.
Note that the curvaton fluctuations can be ignored in two distinct
situations: firstly, if their contribution turns out to be numerically
negligible in the expression for $\zeta$; secondly, if the curvaton
field is massive during inflation (i.e. $m_\sigma\gg H$), in which
case its fluctuations are suppressed.

From the general formula (\ref{zeta_3}), the curvature perturbation in
this case reduces to
\beq
\zeta=-\frac{r}{6}\d_\Gamma
-\frac{r}{72}  \left(r^2+2 r-9\right)\d_\Gamma^2
-\frac{r \left(3 r^4+10 r^3-22 r^2-54 r+135\right)}{1296}\d_\Gamma^3\,.
\eeq
Substituting the expansion of $\d_\Gamma$ in terms of $\delta\chi$,
given in (\ref{expansions}), one obtains, up to third order, the
expression
\begin{eqnarray}
\label{zeta3_mod}
\zeta&=&-\frac{r}{6}\frac{\Gamma'}{\Gamma}\d\mod
-\frac{r}{72}  
\left[6 \frac{\Gamma^{''}}{\Gamma} +\left(r^2+2 r-9\right) \frac{\Gamma'^2}{\Gamma^2} \right] \d\mod^2 
-\frac{r}{1296} \left[36 \frac{\Gamma^{'''}}{\Gamma} 
 +18 \left(r^2+2 r-9\right)
   \frac{ \Gamma' \Gamma''}{\Gamma^2} 
   \right.
\notag \\
&&
\left.\qquad
   +\left(3 r^4+10 r^3-22 r^2-54
   r+135\right) \frac{\Gamma'^3}{\Gamma^3}\right]\d\mod^3\,. 
\end{eqnarray}
As mentioned earlier, our analysis includes the original modulated
reheating scenario if $\sigma$ is identified with the inflaton. One
can indeed check that the above formula, in the limit $r=1$ (since the
inflaton dominates the total energy density), reduces to the
third-order expression obtained for $\zeta$ in the modulated reheating
scenario\cite{Suyama:2007bg}.

Since the expression (\ref{zeta3_mod}) is exactly of the form
(\ref{N_exp}) with $\chi$ as a unique scalar field, one can readily
use the general formulas (\ref{power}-\ref{eq:tau_NL_N}) to determine
the power spectrum and the non-linearity parameters.  The power
spectrum is thus
\begin{equation}
\label{P_zeta}
P_\zeta   
= 
\left(  \frac{r}{6} \right)^2 
\left( \frac{\Gamma'}{\Gamma} \right)^2 P_*\,,
\end{equation}
while we obtain for the non-linearity parameters
\begin{eqnarray}
\tf & = &\frac{3}{r} \left(3-2 \frac{ \Gamma\Gamma ''  }{\Gamma^{'2}}\right) -2-r, \qquad \tau_{\rm NL}=\tf^2\,,
\\ \notag \\
\tg
 & = &  
 \frac{1}{r^2} \left[
 36 \frac{ \Gamma^2 \Gamma^{'''}}{\Gamma'^3} 
 + 18 \left(r^2+2 r-9\right) \frac{ \Gamma  \Gamma''}{\Gamma'^2} +3 r^4+10 r^3-22 r^2-54 r+135
  \right]. \notag \\ 
\end{eqnarray}
In the limit $r=1$, one recovers the predictions of the modulated
reheating scenario. In the limit $r\ll 1$, one finds that the dominant
terms in $f_{\rm NL}$ and $g_{\rm NL}$, barring some special
cancellation, scale like $1/r$ and $1/r^2$, respectively.  This is
different from the standard curvaton model, where both $f_{\rm NL}$
and $g_{\rm NL}$ scale like $1/r$\footnote{
  This is true only if the curvaton potential is quadratic, which we
  assume in the present work.  When the potential is not quadratic,
  $g_{\rm NL} \propto 1/r^2$ \cite{Enqvist:2008gk}.
}. This specific feature of our model can be expressed by the relation
\begin{equation}
g_{\rm NL} =C f_{\rm NL}^2,\qquad (r \ll 1)
\end{equation}
where the numerical value of the coefficient $C$, and in particular
its sign, depends on the functional form of $\Gamma$:
\begin{equation}
C = \frac23 
\left( 15 + 4  \frac{\Gamma^2 \Gamma^{'''}}{\Gamma'^3}  -18 \frac{ \Gamma  \Gamma''}{\Gamma'^2} \right)
\left( - 2 \frac{ \Gamma  \Gamma''}{\Gamma'^2}  + 3 \right)^{-2}\,.
\end{equation}

To go further and determine quantitatively the parameters $f_{\rm NL}$
and $g_{\rm NL}$, taking into account the observed amplitude of the
power spectrum, we need to assume specific expressions for the
function $\Gamma (\mod)$. We consider below three possibilities for
$\Gamma (\mod)$.

\bigskip
\bigskip
\noindent
{\bf Case I}: $\displaystyle \Gamma (\mod)= \Gamma_0\,  \chi^p$ ($p\geq0$).\\

By imposing the CMB normalization for the power spectrum ${\cal
  P}_\zeta = 2.4 \times 10^{-9}$ \cite{Bennett:2012fp},
Eq.~(\ref{P_zeta}) determines the parameter $r$ in terms of the
inflationary Hubble parameter $H_{\rm inf}$ and $\mod$:
\begin{equation}
r = \frac{1.8 \times 10^{-3}}{p} \frac{\mod}{H_{\rm inf}}.
\end{equation}
Furthermore, $r$ should be less than unity by definition, which limits
the parameter range of $H_{\rm inf}$ and $\mod$ to satisfy
\begin{equation}
H_{\rm inf} \gtrsim \frac{4.5 \times 10^{15}~{\rm GeV}}{p} \frac{\mod}{M_{\rm pl}}.
\end{equation}

\begin{figure}[htbp]
  \begin{center}
    \resizebox{180mm}{!}{
     \includegraphics{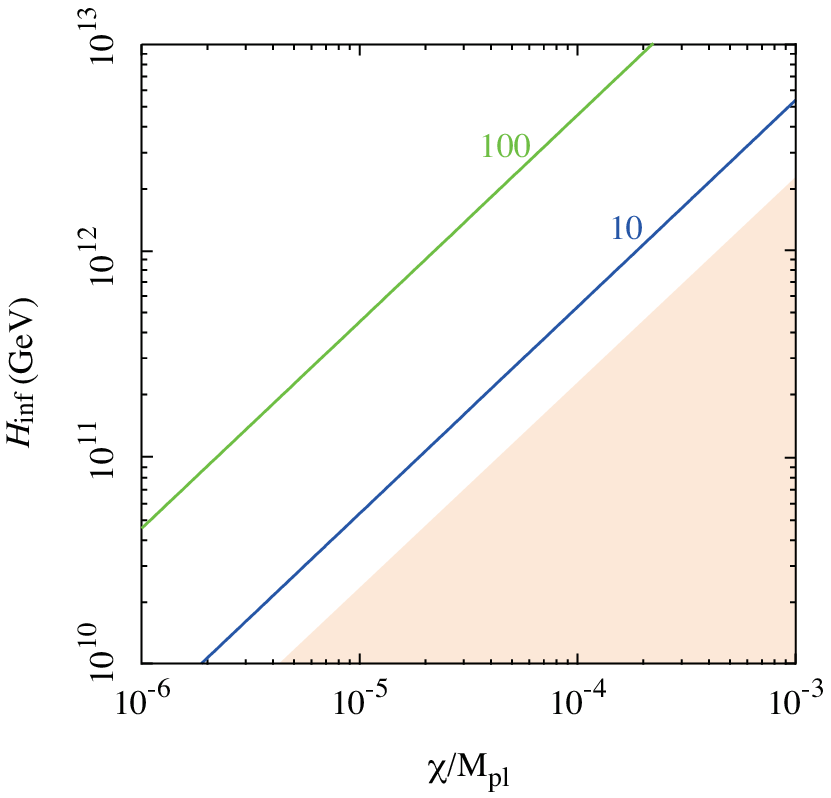}
    \includegraphics{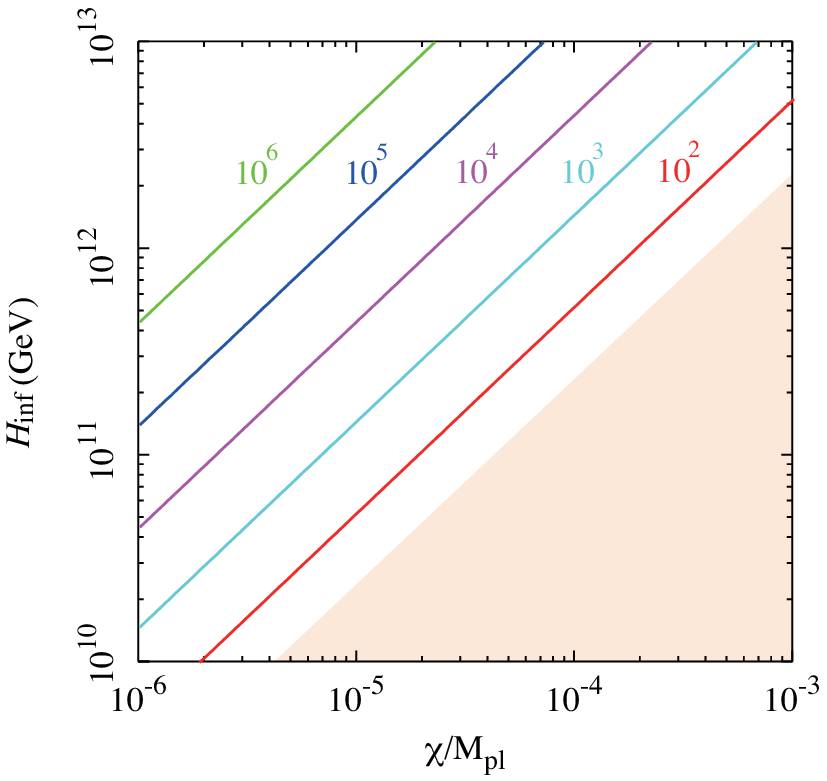}
}
  \end{center}
  \caption{Contours of $f_{\rm NL}$ (left) and $g_{\rm NL}$ (right) in
    the $\chi$--$H_{\rm inf}$ plane for the Case I. $p=2$ is
    assumed. Shaded regions correspond to the parameter space where we
    cannot obtain the right amplitude for $P_\zeta$. }
  \label{fig:fNL}
\end{figure}

Regarding the non-linearity parameters $f_{\rm NL}$ and $g_{\rm NL}$,
the dependence on the functional form of $\Gamma$ only appears in the
combinations $\Gamma \Gamma^{''}/ (\Gamma')^2$ and $\Gamma
\Gamma^{'''}/ (\Gamma')^3$.  By substituting the functional form, we
obtain
 \begin{eqnarray}
\tf & = &  \frac{3 (p+2)}{p\,  r}-2-r,
\\ \notag \\
\tg
 & = &  \frac{p^2 \left(3 r^4+10 r^3-4 r^2-18 r+9\right)-18 p \left(r^2+2 r-3\right)+72}{p^2\, 
   r^2}
 . \notag \\ 
\end{eqnarray}
In the limit $r \ll 1$, one thus finds that $f_{\rm NL}$ is always
positive and can become large if $r$ is small enough. In the same
limit, $g_{\rm NL}$ is positive and enhanced by the factor $1 /r^2$,
as already mentioned.  In Fig.~\ref{fig:fNL}, we show contours of
$f_{\rm NL}$ (left plot) and $g_{\rm NL}$ (right plot) in the
$\mod$--$H_{\rm inf}$ plane. The value of $r$ is fixed by the CMB
normalization as described above.

\bigskip
\bigskip
\noindent {\bf Case II}:
$\displaystyle\Gamma (\mod) = \Gamma_0 \left[ 1+  a \frac{\chi}{M}  +  b \left( \frac{\chi}{M} \right)^2 \right]$. \\

In many models, the coupling can be written as a Taylor expansion of
$\mod$, where $M$ represents some high energy scale and $\chi / M \ll
1$ is assumed.  The coefficients $a$ and $b$ are parameters of order
one, which are supposed to depend on the details of some explicit
model of high energy physics.  The non-linearity parameters are then
given by
\begin{eqnarray}
\tf & = &  -\frac{12b}{a^2r} -r + \frac{9}{r} - 2,
\\
\tg
& = &  
\frac{1}{r^2} \left[ 
\frac{36 b }{a^2}\left(r^2+2 r-9\right)  +3 r^4+10 r^3-22 r^2-54 r+135
  \right].
\end{eqnarray}
where we have used $\mod/M\ll 1$.  As easily read off from the above
expressions, the signs of $f_{\rm NL}$ and $g_{\rm NL}$ can be
positive or negative, depending on $a$ and $b$.  Since $a$ and $b$ are
assumed to be $\mathcal{O}(1)$, the amplitude of $f_{\rm NL}$ and
$g_{\rm NL}$ is mainly controlled by the value of $r$.

\bigskip
\bigskip
\noindent 
{\bf Case III}: $\displaystyle\Gamma (\mod) = \Gamma_0 \left[ 1+  \left( \frac{\chi}{M}\right)^q \right]$ ($q\geq 2$). \\

In the latter case, we consider the possibility that the expansion
starts with a higher order polynomial. Once again, $M$ is some high
energy scale characterizing the underlying physics, and we assume
$\mod / M \ll 1$. The non-linearity parameters are now given by
\begin{eqnarray}
\frac65 f_{\rm NL} & = & -\frac{6}{r} \frac{q-1}{q} \left( \frac{M}{\mod} \right)^q -r+\frac{9}{r}-2,
\\ \notag \\
\frac{54}{25} g_{\rm NL} 
 & = &  
 \frac{1}{r^2} \left[ 
 36  \frac{(q-1)(q-2)}{q^2}\left( \frac{M}{\mod} \right)^{2q} 
 + 18 \left(r^2+2 r-9\right)  \frac{q-1}{q} \left( \frac{M}{\mod} \right)^q \right.  
 \notag \\
&& \left.  +\left(3 r^4+10 r^3-22 r^2-54 r+135 \right)  \right]. 
\end{eqnarray}
Since $\mod / M \ll 1$, one sees that $f_{\rm NL}$ can be large even
with $r=1$, but the sign of $f_{\rm NL}$ is then negative, which is in
contradiction with the current constraints from WMAP
\cite{Bennett:2012fp}.  On the other hand, when $r \ll 1$, $f_{\rm
  NL}$ and $g_{\rm NL}$ can be both positive, which is similar to the
case I.

\section{Hybrid curvaton-modulaton case}
\label{sec:hybrid}

In this section, we consider the general situation where both the
curvaton and modulaton fluctuations contribute to the final density
perturbation. We also take into account the inflaton contribution in
the power spectrum.

The curvature perturbation in this case is given by \eqref{zeta_3},
with the substitutions of (\ref{expansions}). This leads to an
expression of the form (\ref{N_exp}), which contains now three scalar
fields $\phi$, $\curv$ and $\mod$.  It is convenient to introduce two
dimensionless parameters that characterize the relative contributions
of $\sigma$ and $\chi$ to the total power spectrum, defined by
\begin{equation}
\Xi_\sigma \equiv  \frac{N_\sigma^2}{N_\phi^2+N_\sigma^2 + N_\chi^2},
\qquad\qquad
\Xi_\chi \equiv  \frac{N_\chi^2}{N_\phi^2+ N_\sigma^2 + N_\chi^2}.
\end{equation}
The inflaton contribution in the power spectrum is thus given by
$1-\Xi_\curv-\Xi_\mod$.

The bispectrum parameter $\tf$ can then be decomposed into three terms
(the inflaton does not contribute to the non-Gaussianity here)
\begin{eqnarray}
\tf
&=& \Xi_\curv^2\,  \tilde{f}_{\rm NL}^{(\curv^2)} 
+ 2\, \Xi_\curv \, \Xi_\chi\,  \tilde{f}_{\rm NL}^{(\curv\chi)} 
+ \Xi_\chi^2\,  f_{\rm NL}^{(\chi^2)},
\end{eqnarray}
where we have defined
\begin{equation}
\tilde{f}_{\rm NL}^{(\curv^2)} = \frac{N_{\curv\curv}}{N_\curv^2}, 
\qquad
\tilde{f}_{\rm NL}^{(\curv\mod)} = \frac{N_{\curv\mod}}{N_\curv N_\mod}, 
\qquad
\tilde{f}_{\rm NL}^{(\mod^2)} = \frac{N_{\mod\mod}}{N_\mod^2}\,.
\notag
\end{equation}
Their explicit expressions are
\begin{eqnarray}
\tilde{f}_{\rm NL}^{(\curv^2)} 
& = &  \frac{3}{2r} -2 - r,  \\
 \tilde{f}_{\rm NL}^{(\chi^2)} 
& = & 
  -\frac{6}{r} \frac{ \Gamma\Gamma ''  }{\Gamma^{'2}} -r+\frac{9}{r}-2, \\
\tilde{f}_{\rm NL}^{(\curv\chi)} 
& = &-r+\frac{3}{r}-2,
\end{eqnarray}
where the first equation corresponds to the standard curvaton
expression\cite{Lyth:2005fi}, while the second one is the modulaton
contribution calculated in the previous section.  The final one is a
mixed contribution.

Similarly, the trispectrum coefficient $\tg$ can be decomposed into
four terms:
\begin{eqnarray}
\tg
&=&
\Xi_\curv^3\, \tilde{g}_{\rm NL}^{(\curv^3)} 
+ 3\, \Xi_\curv^2 \, \Xi_\chi \,\tilde{g}_{\rm NL}^{(\curv^2\chi)} 
+ 3\, \Xi_\curv\, \Xi_\chi^2\, \tilde{g}_{\rm NL}^{(\curv\chi^2)} 
+ \Xi_\chi^3 \, g_{\rm NL}^{(\chi^3)},
\end{eqnarray}
where  
\begin{equation}
\tilde{g}_{\rm NL}^{(\curv^3)}  = \frac{N_{\curv\curv\curv}}{N_\curv^3}, 
\qquad
\tilde{g}_{\rm NL}^{(\curv^2\mod)}  = \frac{N_{\curv\curv\mod}}{N_\curv^2 N_\mod}, 
\qquad
\tilde{g}_{\rm NL}^{(\curv\mod^2)} = \frac{N_{\mod\mod\curv}}{N_\curv N_\mod^2}, 
\qquad
\tilde{g}_{\rm NL}^{(\mod^3)} = \frac{N_{\mod\mod\mod}}{N_\mod^3}.
\end{equation}
By using \eqref{zeta_3} and \eqref{expansions}, we can explicitly write down these quantities as 
\begin{eqnarray}
\tilde{g}_{\rm NL}^{(\curv^3)} 
& = &  3 r^2+10 r-\frac{9}{r}+\frac{1}{2}, \\
\tilde{g}_{\rm NL}^{(\curv^2\chi)} 
& = & 3 r^2+\frac{9}{2 r^2}+10 r-\frac{15}{r}-\frac{5}{2} , \\
\tilde{g}_{\rm NL}^{(\curv\chi^2)}
& = & 
\frac{1}{r^2} \left(r^2+2 r-3\right) (\left(3 r^2+4 r-9\right) 
+ \frac{6}{r^2}   \left(r^2+2 r-3\right)  \frac{\Gamma\Gamma^{''}}{\Gamma'^2}, \\
\tilde{g}_{\rm NL}^{(\chi^3)}
& = &  \frac{1}{r^2} \left[
 36 \frac{ \Gamma^2 \Gamma^{'''}}{\Gamma'^3}
 + 18 \left(r^2+2 r-9\right) \frac{ \Gamma  \Gamma''}{\Gamma'^2}+\left(3 r^4+10 r^3-22 r^2-54 r+135\right)
  \right],
\end{eqnarray}
where one can identify the usual curvaton
contribution\cite{Sasaki:2006kq}, the pure modulaton contribution
calculated in the previous section, as well as two mixed
curvaton-modulaton contributions.

Finally, the $\tau_{\rm NL}$ coefficient can be decomposed as
\begin{eqnarray}
\tau_{\rm NL}&=& \Xi_\curv^3 \left(\tilde{f}_{\rm NL}^{(\curv^2)}\right)^2+2\, \Xi_\curv^2\, \Xi_\mod \, \tilde{f}_{\rm NL}^{(\curv^2)}\tilde{f}_{\rm NL}^{(\curv\mod)}+ \Xi_\curv\, \Xi_\mod\left(\Xi_\curv+ \Xi_\mod\right) \left(\tilde{f}_{\rm NL}^{(\curv\mod)}\right)^2
\cr
&&
\qquad \qquad\qquad +2\, \Xi_\curv\, \Xi_\mod^2 \, \tilde{f}_{\rm NL}^{(\mod^2)} \tilde{f}_{\rm NL}^{(\curv\mod)}
+\Xi_\mod^3 \left(\tilde{f}_{\rm NL}^{(\mod^2)}\right)^2\,.
\end{eqnarray}

In the $r\ll 1$ limit, one finds that the dominant terms for the
non-Gaussianity coefficients are
\beq
\tf=\frac{1}{2r}\left(3\Xi_\sigma^2+12\Xi_\sigma\Xi_\chi+18\Xi_\chi^2-12\Xi_\chi^2 \frac{\Gamma \Gamma''}{\Gamma'^2}\right)+{\cal O}(1),
\eeq
\beq
\tau_{\rm NL}=
\frac{9}{4r^2} 
\left[ 
\Xi_\sigma^3
+ 8 \Xi_\sigma^2 \Xi_\chi
+ \Xi_\chi^2  \Xi_\sigma 
\left( 28 - 16 \frac{\Gamma \Gamma^{''}}{\Gamma^2} \right)
+ \Xi_\chi^3 
\left( 36 - 48 \frac{\Gamma \Gamma^{''}}{\Gamma^2}  + 16 \frac{\Gamma^2 \Gamma^{'' 2}}{\Gamma^{'4}}
\right)
      \right] + {\cal O}\left({\frac{1}{r}}\right), 
\eeq
\begin{eqnarray}
\tg
&=&\frac{9 \, \Xi _{\chi }}{2r^2}\left[ 
3  \left(6 \Xi _{\sigma } \Xi _{\chi }+\Xi _{\sigma }^2+10 \Xi _{\chi }^2\right)
   -12 \Xi _{\chi } \left(\Xi _{\sigma }+3 \Xi _{\chi }\right)
  \frac{\Gamma \Gamma''}{\Gamma'^2}
  +  8\Xi _{\chi}^2\frac{ \Gamma^2 \Gamma'''}{\Gamma'^3}
\right]+{\cal O}\left({\frac{1}{r}}\right). \notag \\
\end{eqnarray}
Here it is interesting to notice that the leading term of $g_{\rm NL}$
vanishes when $\Xi_\mod =0$. Thus the enhancement by the factor of
$1/r^2$ comes from the modulaton fluctuations, which is absent in the
standard curvaton model.  This property of the trispectrum could be
useful to discriminate this model from other ones.

\section{Summary}
\label{sec:summary}

We have investigated the density perturbations in a scenario based on
the modulated decay of the curvaton, where the curvature perturbation
is generated via the modulation of the curvaton decay rate due to its
dependence on another scalar field, called the modulaton $\mod$. We
have paid special attention to non-Gaussianity, since it is
potentially the best way to discriminate among various scenarios, and
we have computed specifically the non-linearity parameters $f_{\rm
  NL}$ and $g_{\rm NL}$ in this class of models.
  
As discussed in Section~\ref{sec:modulaton_dom}, this model shares
some similarities with the usual curvaton and modulated reheating
models: the size of these parameters are mainly determined by $r$,
which is similar to the curvaton model.  On the other hand, the signs
of these parameters highly depend on the functional form of $\Gamma
(\mod)$, which is the same as usual modulated reheating model.
However, the model also exhibits interesting predictions coming from a
hybrid nature of this model. When the curvaton is a subdominant
component of the Universe at its decay, the non-linearity parameters
are related as $g_{\rm NL} \propto f_{\rm NL}^2$ where the
proportionality factor depends on the functional form of $\Gamma
(\chi)$.  Thus this model generically predicts enhanced $g_{\rm NL}$,
which is different from the standard curvaton and modulated reheating
models, predicting $g_{\rm NL} \sim f_{\rm NL}$.

We have also investigated a more general case where fluctuations of
the curvaton itself also contribute to density fluctuations. We have
presented the formulas of the curvature perturbation up to the 3rd
order, and the non-linearity parameters.  The formulas given in such a
case naturally also include the standard curvaton and modulated
reheating models, which provides a unified treatment of those kind of
models and their variants.

Although we have considered only adiabatic perturbations in the
present work, the modulated decay of the curvaton could also produce
isocurvature perturbations, similarly to the standard curvaton
scenario\cite{Lyth:2002my}.  It would be interesting to study these
possible isocurvature modes and their non-Gaussianities following the
analysis introduced in \cite{Langlois:2011zz,Langlois:2010fe}.
Isocurvature modes, possibly correlated with adiabatic modes, lead to
very specific signatures in the CMB non-Gaussianities and Planck or
future CMB data could detect these isocurvature
non-Gaussianities\cite{Langlois:2011hn,Langlois:2012tm}\footnote{ For
  constraints on isocurvature non-Gaussianities from current data, see
  \cite{Hikage:2008sk,Hikage:2012be,Hikage:2012tf}.  }.

In the near future, one can hope that new cosmological data of
unprecedented precision, in particular from Planck, will enable to
test the model presented here, together with various other mechanisms
that generate primordial density perturbations.  In this respect, the
unified treatment proposed in this paper should be useful for a
simplified confrontation of a large class of models with cosmological
data.

\bigskip
\bigskip
{\sl Note added:}
While completing this manuscript, we became aware  that the authors of 
\cite{Assadullahi_etal} were  working on a very similar topic.

\section*{Acknowledgments}
T.T would like to thank APC for the hospitality during the visit,
where part of this work has been done. The authors would also like to thank the Yukawa Institute for Theoretical Physics at Kyoto University, 
where this work was completed during the Long-term Workshop YITP-T-12-03 on ``Gravity and Cosmology 2012". 
D.L. is partly supported by the ANR (Agence Nationale de la Recherche) grant ÒSTR-COSMOÓ ANR-09-BLAN-0157-01.
 The work of T.T. is partially
supported by the Grant-in-Aid for Scientific research from the
Ministry of Education, Science, Sports, and Culture, Japan,
No.~23740195.

\end{document}